\documentclass[superscriptaddress,prl,preprint]{revtex4}

\usepackage{epsfig}
\usepackage{amsmath}
\usepackage{graphicx}
\usepackage{dcolumn}
\usepackage{amsmath}
\usepackage{latexsym}
\usepackage{epsfig}
\usepackage{amsmath}
\usepackage{graphicx}
\usepackage{dcolumn}
\usepackage{amsmath}
\usepackage{latexsym}
\usepackage{color}
\usepackage{epstopdf}
\usepackage{subfigure}

\usepackage[ulem=normalem]{changes}

\begin{document}

\title{The effect of Coulomb interactions on the disorder driven superconductor-insulator transition:

THz versus tunneling spectroscopy }
\author{Daniel Sherman}
\affiliation{1.~Physikalisches Institut, Universit{\"a}t Stuttgart, Pfaffenwaldring 57, 70550 Stuttgart, Germany}
\affiliation{Department of Physics, Bar Ilan University, Ramat Gan 52900, Israel}
\author{Boris Gorshunov}
\affiliation{1.~Physikalisches Institut, Universit{\"a}t Stuttgart, Pfaffenwaldring 57, 70550 Stuttgart, Germany}
\affiliation{Prokhorov Institute of General Physics, Russian Academy of Sciences, Vavilov Street 38, 119991 Moscow, Russia}
\affiliation{Moscow Institute of Physics and Technology, 141700, Dolgoprudny, Moscow Region, Russia}
\author{Shachaf Poran}
\affiliation{Department of Physics, Bar Ilan University, Ramat Gan 52900, Israel}
\author{John Jesudasan}
\affiliation{Tata Institute of Fundamental Research, Homi Bhabha Road, Colaba, Mumbai 400005, India}
\author{Pratap Raychaudhuri }
\affiliation{Tata Institute of Fundamental Research, Homi Bhabha Road, Colaba, Mumbai 400005, India}
\author{Nandini Trivedi }
\affiliation{Department of Physics, The Ohio State University, 191 W. Woodruff Avenue, Columbus, Ohio 43210, USA}
\author{Martin Dressel}
\affiliation{1.~Physikalisches Institut, Universit{\"a}t Stuttgart, Pfaffenwaldring 57, 70550 Stuttgart, Germany}
\author{Aviad Frydman}
\affiliation{Department of Physics, Bar Ilan University, Ramat Gan 52900, Israel}

\maketitle

\textbf{The interplay between disorder and superconductivity has intrigued physicists for decades.  Of particular interest is the influence of disorder on the superconducting energy gap $\Delta$. In the absence of Coulomb interactions  between electrons, disorder leads to emergent granularity of the local order parameter resulting in a pseudogap at temperatures above the critical temperature $T_c$, as well as
a finite gap $\Delta$ on the insulating side of the disorder-driven superconductor-insulator transition (SIT). At the same time, disorder also enhances the Coulomb interactions, which subsequently may influence $\Delta$ in a manner that is still not fully understood. Here we investigate the evolution of the energy gap through the SIT by two different experimental methods: tunneling spectroscopy, in which a metallic electrode is placed close to the studied sample thus screening the Coulomb interactions, and terahertz (THz) spectroscopy, which probes the unscreened sample. The comparison between the two methods  illustrates the role played by electronic interactions in determining the nature of the phases across the SIT and sheds light on the mechanisms involved in the destruction of superconductivity. }

\vspace{1cm}

\noindent \textbf{Introduction.} The effect of disorder on superconductivity, even for s-wave superconductors, is an outstanding problem, yet unresolved. It involves the interplay of three major phenomena: superconductivity describing the pairing of electrons due to retarded attractive interactions, localization of wave functions because of strong disorder, and the evolution of Coulomb interactions with increasing disorder. It was argued by Anderson \cite {anderson} that superconductivity is not affected by disorder on the scale of the superconducting energy gap $\hbar/\tau \approx \Delta$, where $\tau^{-1}$ is the scattering rate, persisting even in polycrystalline or amorphous materials. Experimentally it was found that superconductivity can be destroyed by a sufficiently large degree of  disorder of the order of the Fermi energy \cite{strongin, goldman1, dynes1,dynes2,yazdani,goldman2} $\hbar/\tau \approx \epsilon_F$. Once superconductivity is destroyed the sample undergoes a transition to an insulating state across a superconductor insulator transition (SIT), a fundamental manifestation of a quantum phase transition at $T=0$ tuned by a non-thermal parameter such as disorder or a magnetic field.

In this paper we focus on the effect of disorder on the energy gap $\Delta$. According to the BCS theory the relationship between the critical temperature $T_c$ and the energy gap $\Delta$ is given by $ 2\Delta/{k_B T_c}\equiv \alpha$
where the weak coupling BCS result gives $\alpha \approx 3.5$. The Mcmillan formula \cite{mcmillen} in the strong electron-phonon coupling limit provides no such explicit formula but experimentally $\alpha$ is found to increase with increasing coupling strength reaching a typical value of 4.8 for strong coupling superconductors such as Pb \cite{allen}.

Disorder may influence the energy gap in various ways. Finkelstein \cite{finkelstein} showed that in uniform superconductors, disorder induces diffusive transport which in turn intensifies Coulomb repulsion between electrons. This suppresses electron pairing leading to a decrease of both the critical temperature and the energy gap with increasing disorder. Indeed a thin normal metal was shown to increase $T_c$ and $\Delta$ of a nearby disordered superconductor by screening the coulomb interactions \cite{bourgeois}.

Another possibility is that the superconductor, which  exhibits a many-body coherent state of pairs on the scale of the coherence length $\xi\gg k_F^{-1}$  ($k_F$ is the Fermi wave vector)
becomes spatially inhomogeneous~\cite{kowal} with increasing disorder. This gives rise to an emergent granularity of the pairing amplitude on the scale of $\xi$ even
when the random potential fluctuates on an atomic scale~\cite{ghosal1,ghosal2,feigelman}.
It has been further shown that phase fluctuations riding on top of the
inhomogeneous pairing amplitude suppress the superfluid stiffness $D_s$ at a critical disorder, signaling the superconductor to insulator transition~\cite{Nandini}.
However, the energy gap remains robust and finite
across the transition and also survives as a pseudogap above the critical temperature $T_c$. Indeed, recent tunneling measurements on InO and TiN films have confirmed these theoretical predictions and showed that as disorder increases and the sample approaches the SIT, $\alpha$ increases monotonically and ultimately even diverges at the SIT where $T_c \rightarrow 0$ but the energy gap $\Delta$ is still finite. Consequently, a finite pseudogap was observed above the critical temperature \cite{sacepe,sacepe2,pratap} and a hard gap beyond the
critical disorder in the insulating phase \cite{sherman}.

An important question in the field remains open: how does the combination of both disorder induced ``granularity" and strong Coulomb interactions affect the energy gap?
Does the joint contribution lead to a constant $\alpha$ or a change of $\alpha$ with disorder? We employ two different experimental methods to probe disordered superconductors and compare the results of the energy gap:
The first is tunneling spectroscopy which requires a close metallic electrode that naturally induces screening and thus suppresses Coulomb interactions.
The second is THz spectroscopy which utilizes free space radiation through the sample, with no metallic electrodes or contacts necessary thus enabling the investigation of the unscreened unperturbed sample.
As a cross-check we also perform THz spectroscopy measurements in the presence of a metal electrode. Our main observations are:
(1) When the Coulomb interactions are screened there is a finite gap in both the superconductor and the insulator, albeit with reduced coherence peaks in the insulator.
(2) When the Coulomb interactions are unscreened the gap in the disordered superconductor is substantially suppressed and in the insulator no discernible gap is detected. Furthermore, the gap value $\Delta$ is suppressed with disorder much faster than $T_c$ and consequently  $\alpha$ decreases monotonically to 0 as the sample is driven through the SIT.
\vspace{1cm}

\begin{figure}
\vspace{0cm}
\includegraphics[width=1\textwidth]{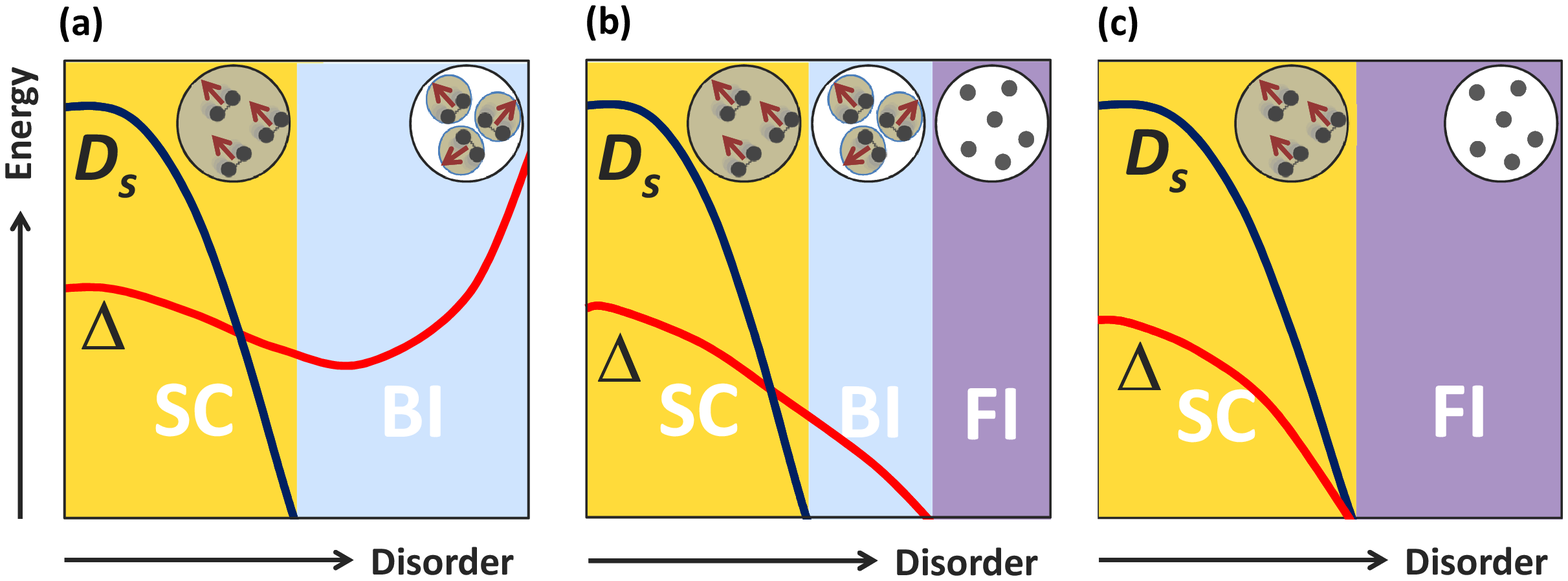}
 \vspace{-8cm}
\caption{\textbf{Paradigms for the disorder driven superconductor-insulator transition (SIT).} Schematic description of three paradigms (see text) for the effect of disorder on the characteristic energy scales of a superconductor: the energy gap $\Delta$ (red line) and the superfluid stiffness $D_s$ (dark blue line).  The interplay between these energies defines three phases: a superconductor (SC) containing free cooper pairs, a Bose insulator (BI) containing localized pairs and a Fermi insulator (FI) containing localized quasi-particles.}
\label{paradigms}
\end{figure}

\noindent \textbf{Paradigms for the SIT.} The disorder driven SIT can be conceptualized in terms of three paradigms depicted in Fig.~\ref{paradigms}. These are
based on the behaviour of energy scales obtained from the interplay of BCS superconductivity, Anderson localization and Coulomb correlations.
Paradigm (a) shows the behaviour of the energy gap $\Delta$ and the superfluid stiffness $D_s$ with increasing disorder, based on calculations that include inhomogeneous
pairing amplitude and phase fluctuations~\cite{ghosal1,ghosal2,Nandini} but ignore Coulomb correlations between the electrons. $D_s=(\hbar^2/m) n_s$ (where m is the Cooper pair mass and $n_s$ is the superfluid density) has dimensions of energy for two-dimensional superconductors and can be interpreted as the energy associated with phase fluctuations.
$T_c$ is determined by the minimum of the two energy scales. The vanishing of $D_s$ at a critical disorder signals a quantum phase transition from a superconductor to a gapped Bose insulator, where Cooper pairing still survives in the insulator in the form of localized Cooper pairs \cite{fisher}, as also confirmed by tunneling experiments
\cite{sacepe,sacepe2,pratap,sherman}.

Panels (b) and (c) illustrate the additional effects of Coulomb interactions and screening, or rather the lack of it, with increasing disorder.
Depending on the interplay between $\Delta$ and $D_s$ there are two possibilities that govern the nature
of the phases and the transition. In paradigm (b),
there is a large regime where $\Delta$ in the presence of both disorder and Coulomb interactions
is finite. Once $D_s$ dips below $\Delta$, the transition temperature $T_c$  is determined by $D_s$, the minimum of the two scales.
This results in a transition from a superconductor to a Bose insulator driven by
vanishing $D_s$ and a further crossover from a Bose-insulator of localized Cooper pairs
to a Fermi insulator of broken pairs.
This paradigm may be relevant for the observed giant magnetoresistance peak in several disordered superconductors \cite{hebard,gantmakher,sanquer,murty,kapitulnik,baturina,dubi,armitage,valles}.

In paradigm (c) enhanced Coulomb interactions suppress $\Delta$
to zero and consequently also $D_s$ which vanishes at the same critical disorder. In this case one can expect a transition from a superconductor to a Fermi insulator directly;
this indeed appears to be the case in the present THz results presented below.

We next discuss our experimental results and return to a discussion of what they imply in the context of the above paradigms.

\vspace{1cm}

\noindent \textbf{Experimental results and discussion.}
We have employed tunneling spectroscopy and THz optical measurements as the prime methods to determine the energy gap of disordered superconductors. To that end, the tunneling spectra obtained at certain temperatures are fitted by the BCS density of states; the complex THz conductivity is described by the BCS-based Mattis-Bardeen theory \cite{MattisBardeen,Zimmermann91} (see methods section).
Superconductivity is reflected in the characteristic behaviour of the complex conductivity $(\sigma_1 + i\sigma_2)$ as a function of photon energy $h f$, where  $f$ the frequency  and $h$ Planck's constant.
 For frequencies $f = 2\Delta/ h$ a kink in $\sigma_1(f)$ is observed due to the onset of radiation absorption; $\sigma_2$ diverges as $1/f$ indicating the presence of a finite superfluid stiffness $D_s$. Hence this method has the additional value that it allows the extraction of  both $\Delta$ and $D_s$. We have utilized these two techniques to determine the behaviour of $\Delta$ through the SIT in two disordered superconducting materials: amorphous indium oxide (InO) films and niobium nitride (NbN) films. In both materials the chemical composition can be manipulated thus enabling preparation of batches of samples with different disorder strengths, characterized by the normal state sheet resistance $R_\Box$ of the films spanning the SIT.
 In both systems a pseudogap has previously been observed in the tunneling DOS  for $T>T_c$ \cite{sacepe2,pratap}. In InO a similar gap was also observed for the insulating phase of the SIT \cite{sherman}.

\begin{figure}[h]
\vspace{0cm}
\includegraphics[width=0.8\textwidth]{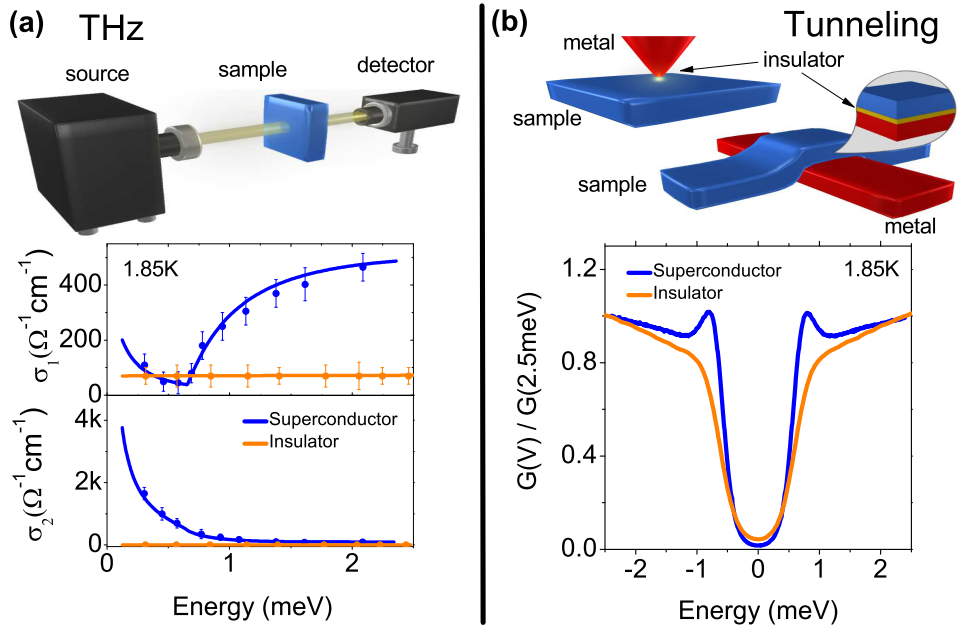}
 \vspace{0cm}
\caption{\textbf{THz versus tunneling spectroscopy on insulating and superconducting indium oxide films. $T_c=3.3K$.   a}, The real part of the conductivity (upper panel) and the imaginary part of the conductivity (lower panel) obtained by using two different methods:  BCS model (solid curves) and a manual calculation based on Fresnel equations (dots with error bars), see methods section for more details. Note that while the superconducting film exhibits a Mattis-Bardeen behaviour typical for superconductors, the insulating film exhibits constant behaviour typical for normal state.  \textbf{b}, Tunneling results on similar indium oxide films. Both the insulating and the superconducting films exhibit a similar energy gap \cite{sherman}, the surrounding depleting background indicates the degree of localization \cite{AA}.}
\label{thz_tun}
\end{figure}

Figure \ref{thz_tun} compares the THz and tunneling results for two InO samples close to the SIT, one on the superconducting side and the other on the insulating side.
The THz measurements reveal a finite gap ($2\Delta=0.62$ meV) and finite superfluid stiffness ($D_s=7$~meV) in the superconducting phase but there is no sign of a gap nor for superfluid stiffness down to $\sim 0.05$~meV for the insulator. On the other hand, tunneling measurements conducted on the same two samples exhibit a finite gap on both sides of the transition. Though neither curves fit the BCS tunneling DOS very well, the best fits yield $\Delta\approx 0.7$~meV for both the superconductor and the insulator (see ref. \cite{sherman}).
Similar trends were observed in previous planar junctions and STM measurements on both InO and NbN films \cite{sacepe2, pratap}.
Evidently, the presence of a tunneling electrode not only enhances the gap in the superconducting phase, but in addition, the gap remains finite in the insulating phase.

\begin{figure}[h]
\vspace{3cm}
\includegraphics[width=1\textwidth]{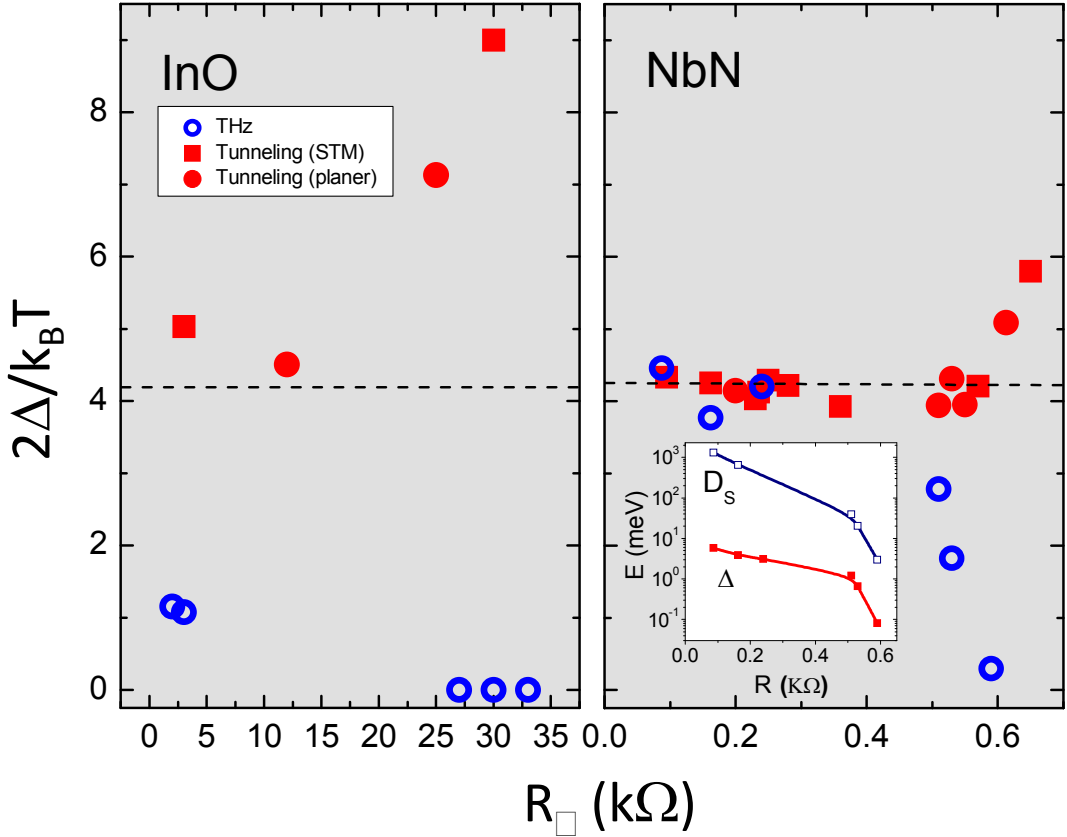}
 \vspace{-5cm}
\caption{\textbf{THz versus tunneling spectroscopy for series of NbN and InO films with different disorder.} $\alpha\equiv 2\Delta /k_B T_c$ as a function of sheet resistance for batches of InO films (left) and NbN films (right) obtained by planar junction tunneling (red squares), STM tunneling (red circles), and THz spectroscopy (blue circles). The dashed line indicates the experimental value of clean samples. Some of the tunneling results were taken from \cite{sacepe2} and \cite{pratap}. Note that the results from planar junction and STM methods agree with each-other very well. Inset: The energy gap and the superfluid stiffness as a function of sheet resistance for the NbN batch. The resemblance to paradigm c (Fig. \ref{paradigms}c) is evident.}
\label{master}
\end{figure}

This trend is further demonstrated in Fig. \ref{master} which depicts $2\Delta/{k_B T_c}$ obtained by tunneling and by THz spectroscopy in batches of NbN and InO films having different degrees of disorder. For relatively clean samples, both methods yield the same $\Delta$ and correspondingly the same $\alpha=2\Delta/{k_B T_c}$. As disorder becomes stronger, $T_c$ decreases and eventually approaches zero, however, tunneling measurements always observe a finite gap, resulting in an increase of $\alpha$ with disorder. The results obtained from THz spectroscopy are quite the opposite. With increasing disorder, an energy gap can still be extracted from the conductivity data yielding good fits to BCS theory; however its energy scale is significantly suppressed compared to the BCS expectations. In both InO and NbN films $\alpha$  \emph{decreases} rapidly  with disorder in stark contrast to the trend observed in tunneling experiments. In fact, $\Delta$ becomes unmeasurable in the THz frequency range well before $T_c$ reaches $0$.

The fact that the THz and tunneling spectroscopy agree very well for clean films but show opposite trends in the strongly disordered limit demonstrates the crucial role played by Coulomb interactions.  The tunneling electrode influences the measurement by screening long range interactions, thus probing a sample in which interactions are fully or partially ``switched off". In this limit $\alpha$ increases with disorder.
The THz experiment measures an unperturbed sample with Coulomb interactions present. The "unscreened" $\Delta_{THz}$ is clearly smaller than the ``screened" $\Delta_{\rm tunneling}$ and the difference between the two becomes larger with increasing disorder.

The inset of Fig. \ref{master} depicts the two energies $\Delta$ and $D_s$, extracted from the THz data, as a function of disorder represented by $R_\Box$. Both scales drop sharply to zero with increasing disorder in a way that clearly resembles Fig. \ref{paradigms}c. This behavior of the gap and the superfluid stiffness support the notion that the samples experience a transition from a superconductor directly to a Fermi insulator.

\begin{figure}[h]
\includegraphics[width=0.7\textwidth]{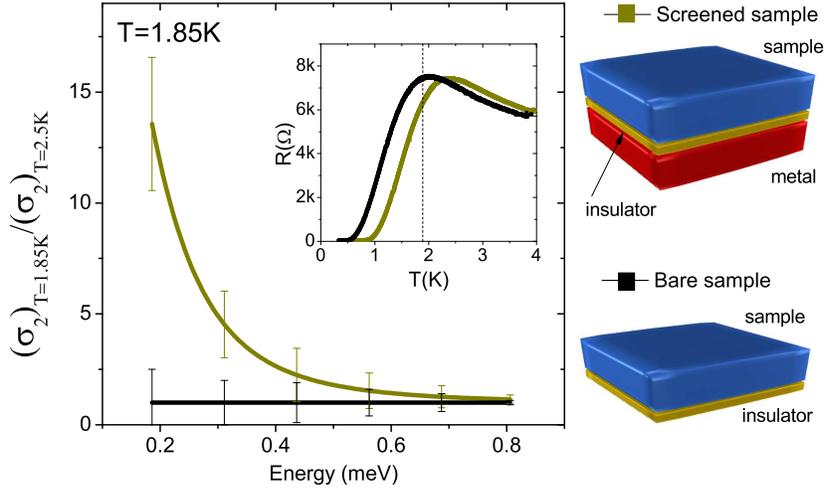}
   \vspace{-0.5cm}
 \caption{\textbf{THz spectroscopy of an indium oxide samples with and without a nearby metal plane.} The imaginary part of the conductivity for the bare sample (black curve) and the screened sample (dark yellow curve) obtained at $T=1.85K$ and normalized to the results at $T=2.5K$. The rapid increase in the imaginary part, seen only in the screened sample, indicates the superconducting DC conductivity by virtue of Kramers-Kronig relation. The bare sample shows constant behaviour as expected from the normal state. The difference between the two identical InO samples is further illustrated in the inset which shows the resistance versus temperature for both samples. The screened sample exhibits a higher critical temperature than the bare sample.}
 \label{screen}
 \end{figure}

Next we present results from the cross-check experiment in which the gap is determined from the THz experiment on films with and without a nearby metal plate.
For this purpose we fabricate two InO films which are deposited together under identical conditions, one is a ``bare" sample on an insulating substrate and the other
a ``screened" sample separated from an Al metal layer by a 5~nm thick insulator. As seen in the inset of Fig. \ref{screen}, the presence of the metal plate increases the $T_c$ of the film. $\sigma_2$ measured at $T_c^{\rm bare} <T < T_c^{\rm screened}$ for both samples is shown in Fig. \ref{screen}. While the bare sample shows no change in $\sigma_2(f)$ and remains constant at these energies as expected for the normal state, the screened sample exhibits a divergence of $\sigma_2$ at low $f$ indicating the presence of a non-zero $D_s$ and a finite superconducting gap $\Delta\sim 0.1$~meV. Note that in this setup the direct measurement of $\sigma_1$ is less accurate due to the absorption of the metal film which drastically cuts down the transmission; nevertheless, the phase shift and consequently $\sigma_2$ are measured reliably. From the above results we make two important conclusions: First, unlike tunneling results, THz measurements show no gap for $T>T_c$ and secondly, at $T=1.85$K the presence of the nearby metallic plane induces a gap in the film that is otherwise insulating and gapless, thus revealing the role of Coulomb interactions in suppressing $\Delta$.

 Still, the fact that in the presence of interactions disorder affects $\Delta$ much more than $T_c$ is puzzling. According to Fig.~\ref{paradigms}(c) $T_c$ should
 be determined by $\Delta$, the smaller of the two energy scales. But why does $\alpha$ decrease significantly with disorder? The
prediction of Finkelstein \cite{finkelstein} captures the suppression of the pairing gap due to enhanced Coulomb interactions. However, it does not explain why
large disorder suppresses the gap much stronger than $T_c$.
While a detailed theory still needs to be fleshed out one may consider the influence of the normal state DOS. Coulomb interactions play a significant role in restructuring the DOS of a disordered normal metal close to the metal-insulator transition.
For weakly localized metallic films the interplay between Coulomb interactions and disorder leads to a depletion of the DOS  around the Fermi level \cite{AA} in what is known as the zero bias anomaly.
For strongly localized systems, electron interactions open a soft Coulomb gap at the Fermi energy as shown by Efros and Shklovskii \cite{ES}. Indeed, all tunneling measurements on disordered superconductors exhibit a normal-state background showing suppression of the density of state around zero energy \cite{pratap, sherman}. Hence, the superconducting energy gap must open up on this background of a strongly depleted density of state. It is reasonable that this may have a considerable effect on the superconducting gap energy. The THz experiments imply that this is indeed the case and we expect them to invoke theoretical ideas on the interplay between disorder and interactions that would shed light on the observed results.

\vspace{1cm}

\textbf{Conclusions} The set of results presented here reveal the important role Coulomb correlations play on the energy gap of disordered superconductors. In the absence of Coulomb interactions, disorder suppresses $D_s$ and hence $T_c$ much more than the gap. This is indeed what is observed by the tunneling measurements (in which interactions are naturally screened due to the presence of a metallic electrode or a tip) where a gap is observed for $T>T_c$ as well as in the insulator. Thus the tunneling measurements are well described by paradigm (a). Unscreened Coulomb interactions produce the opposite trend as revealed by THz results. The surprising finding is that $\alpha=2\Delta/ K_BT_c$ decreases monotonically with increasing disorder indicating that $\Delta$ is suppressed by disorder more than $T_c$. This can not be accounted for by the existing paradigms for the SIT. The results also show that when Coulomb interactions are screened by placing a metal electrode a gap is recovered in the insulator. This is a very significant result suggesting the possibility of tuning a quantum phase transition from a gapless Fermi insulator to a Bose insulator by enhanced screening of the Coulomb interactions. Further theoretical studies that incorporate the effects of both the emergent granularity and Coulomb interactions are essential for explaining our experimental results and for a better understanding of when the different paradigms discussed above are borne out in experiments.

\vspace{2cm}

\textbf{Methods}

The InO films were deposited on $10$x$10 mm^2$ of THz-transparent MgO or sapphire substrates (with various thickness ranging from 0.5 to 1.5 mm) by e-gun evaporation. During the deposition process dry oxygen was injected into the chamber; the partial oxygen pressure allows us to tune the disorder. The NbN films were grown on similar MgO substrates by reactive magnetron sputtering, where the Nb/N ratio in the plasma served as a disorder tuning parameter. In both cases the deposited films were amorphous yet structurally homogeneous; the thickness ranges from 15 to 40~nm. DC transport measurements were used to characterize $T_c$.

For the samples with a screening metal plane (Fig. \ref{screen}) we used the following procedure: A 30nm thick Al film was deposited on an insulating substrate. The Al was allowed to oxidize in ambient conditions for 24 hours. This process has been shown to produce $\sim 5nm$ thick pinhole free high quality AlO insulators \cite{sherman}. The $T_c$ of the Al layer was $1.1K$, well below the measurement temperature. A disordered SC film was then deposited on the AlO film.  As a cross-check, three additional pairs of "screened" and "bare" samples were produced on different substrates (SiO, sapphire or MgO) all showing similar differences in $T_c$. In particular a pair of samples evaporated on the same substrate, $Si+Ag+AlO$, having different thickness of the AlO insulator exhibited a lower $T_C$ for the sample having a thicker insulator.

THz spectroscopy was used in the past to confirm the BCS theory since it probes the energy range of the superconducting gap \cite{tinkham1,tinkham2,Kornelsen91}. Optical methods offer a number of crucial advantages over DC and tunneling measurements. In the first place a macroscopic area is irradiated homogeneously through the whole sample; thus it is a true volume averaging measurement unlike DC methods that are more sensitive to actual current trajectories and film morphology. Secondly it relinquishes the need for metal contacts which may affect the electron system by screening. Finally, THz spectroscopy can be employed to study also samples very deep in the insulating phase of the SIT, which are not accessible by tunneling methods.
The experimental setup \cite{Gorshunov05} is based on backward wave oscillators as
powerful radiation sources to emit continuous-wave, coherent radiation which can be tuned over the frequency range of $0.05 - 1.1$~THz, corresponding a photon energy of $0.18 - 4.5$~meV.  The beam traversing through the sample is subject to multi-reflection on the two parallel boundaries of the substrate (i.e. Fabry-Perot interference) which
strongly increases the interaction with the film. Subsequently it is detected by a Golay cell or He-cooled bolometer. A reference arm completes the arrangement for a Mach-Zehnder interferometer in order to measure the phase shift in addition to the sample transmission for each frequency. The specimens were mounted in an optical $^4$He bath cryostat, allowing to cool down to $T=1.85$~K.

\begin{figure}
\vspace{1cm}
\includegraphics[width=1\textwidth]{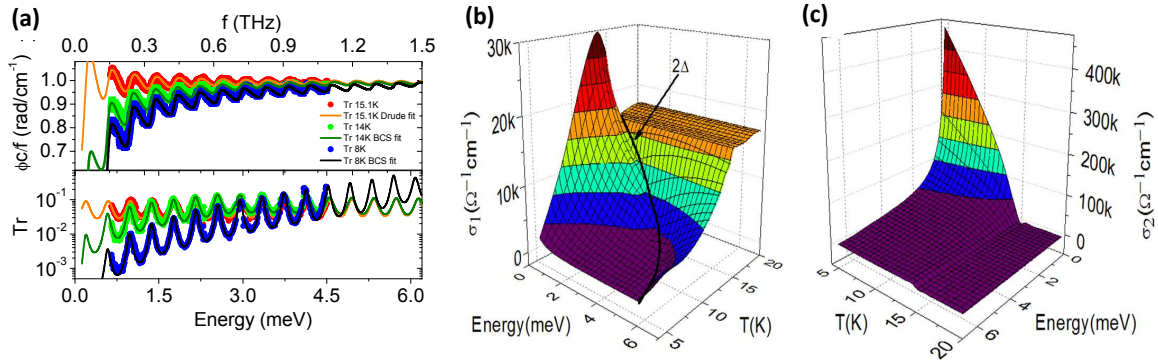}
   \vspace{-8cm}
 \caption{\textbf{Obtaining the energy gap value and the complex conductivity from the transmission and phase-shift data.} \textbf{a}, $\phi(f)c/f$ and Transmission versus energy (frequency) of a NbN film having a critical temperature of 15~K.  The thick curves are the raw data and the thin curves are the fits to Mattis-Bardeen. The growing deviation of the phase shift over frequency from the normal state at $T<T_{c}$ with lowering $T$ or $f$ indicates the increasing superfluid stiffness. The peak in the transmission is a precursor for the value of $2\Delta$.  \textbf{b}, Real part of the complex conductivity $(\sigma_1)$ as a function of temperature and energy. \textbf{c}, Imaginary part of the complex conductivity $(\sigma_2)$ as a function of temperature and energy. }
 \label{method}
 \end{figure}

For each temperature and applied frequency we measured simultaneously the transmission, $Tr$, and phase shift, $\phi$, of the radiation passing through the studied sample, i.e. the film on the substrate; the bare substrate was characterized beforehand. Obtaining these two independent parameters enabled us to directly extract the complex response function (e.g. the real and imaginary parts of the complex conductivity, $\sigma_1$ and $\sigma_2$, respectively) as demonstrated in Fig. \ref{method} \cite{dressel2}.

There are two ways to determine the energy gap $\Delta$ from the raw data.
In the first approach, $Tr(f)$ and $\phi(f)$ are fitted by Fresnel's equations for a thin film on a transparent substrate \cite{dressel1} with frequency dependent optical parameters derived from microscopic models.
In the normal state  the maximum of the Fabry-Perot oscillations hardly vary with frequency; the conductivity is given by the Drude model with the DC conductivity and the scattering rate as fitting parameters.
For $T< T_c$ both $Tr$ and $\phi$ change dramatically with $f$. In this case the THz data are fitted to the Mattis-Bardeen equations complemented by a finite scattering rate \cite{Zimmermann91}; the gap $2\Delta$ and the plasma frequency $\omega_p$ are the fitting parameters. This procedure is demonstrated in Fig. \ref{method}a for a film of NbN.

The second analysis procedure does not rely on any microscopic model.
From the two independent quantities $Tr$ and $\phi$ the complex conductivity, $\sigma_1$ and $\sigma_2$, is calculated via the Fresnel equations (for multiple reflections) \cite{dressel1}. From the frequency dependent conductivity, the value of $2\Delta$ can be easily extracted  by the kink in $\sigma_1(f)$.
Photon energies $h f > 2\Delta$ allow excitations across the superconducting gap, leading to a sharp increase in absorption. Radiation with smaller frequency ($hf < 2\Delta$) can not break Cooper pairs, and only normal quasi-particle absorption is possible.
With lowering the temperature, quasi-particle excitations become sparse and  $\sigma_1(f)$ remains flat down to lowest frequencies or increase with a further decrease in frequency - depending on the value of the scattering rate.
$\sigma_2$ (which mainly depends on $\phi/f$) diverges as $f \rightarrow 0$; according to the Kramers-Kronig relation, this corresponds to a strong zero frequency peak, representing superconductivity. This analysis procedure was employed to produce the plots in Fig. \ref{method} panels b and c.

\vspace {1cm}

\textbf{Acknowledgements:} We are grateful for  useful discussions with Misha Reznikov and for assistance from Uwe Pracht and Eli Farber.  We acknowledge support by the Deutsche Forschungsgemeinschaft (DFG) and by the US Israel binational fund (grant No. 2008299).

\end{document}